\documentclass[traditabstract]{aa}     

\makeatletter 
\AtBeginDocument{\let\linenumbers\relax}
\makeatother

\usepackage{natbib}
\usepackage{graphicx}
\usepackage{txfonts}
\usepackage{xcolor}
\usepackage{lipsum}
\usepackage{subcaption}         
\usepackage{lscape}             
\usepackage{placeins}           
\usepackage{hyperref}

\newcommand{\bd}{\begin{displaymath}}
\newcommand{\ed}{\end{displaymath}}
\newcommand{\be}{\begin{equation}}
\newcommand{\ee}{\end{equation}}
\newcommand{\beaa}{\begin{eqnarray*}}
\newcommand{\eeaa}{\end{eqnarray*}}
\newcommand{\bea}{\begin{eqnarray}}
\newcommand{\eea}{\end{eqnarray}}
\usepackage{orcidlink}

\def\HST{{\it HST}\xspace}

\def\GLEE{\textsc{Glee}\xspace}

\def\S24{\texttt{S24}\xspace}
\def\Sw{\texttt{Sw}\xspace}
\def\Ew{\texttt{Ew}\xspace}
\def\Swo{\texttt{Swo}\xspace}
\def\Ewo{\texttt{Ewo}\xspace}

\def\Ho{\ensuremath{66.0 ^{+ 1.6 }_{ -1.8 }}\xspace}

\def\HoHoom{\ensuremath{65.0 ^{+ 2.0 }_{ -1.8 }}\xspace}
\def\omHoom{\ensuremath{0.30 ^{+ 0.08 }_{ -0.09 }}\xspace}
\def\f{2.0\xspace} 
\def\relerr{3.0\%\xspace} 

\def\kmsMpc{\xspace\ensuremath{ \text{~km~s}^{-1}\, \text{Mpc}^{-1}}\xspace}

\def\gleetoolspy{\textit{glee}$ \_ $\textit{tools.py} \xspace}

\begin{document}

   \title{Narrowing down the Hubble constant value by exploiting the extended surface brightness of the supernova Refsdal host galaxy}

   \titlerunning{Narrowing down the Hubble constant value with SN Refsdal}


   \author{S.~Schuldt\inst{1}\inst{,2}\inst{,3}\inst{,4\orcidlink{0000-0003-2497-6334}}
   \and C.~Grillo\inst{3}\inst{,4\orcidlink{0000-0002-5926-7143}}
   \and {S.~H.~Suyu}\inst{5}\inst{,6,\orcidlink{0000-0001-5568-6052}}
   \and A.~Acebron\inst{7}\inst{,4,\orcidlink{0000-0003-3108-9039}}
   \and L.~Bazzanini\inst{8}\inst{,9}
   \and P.~Bergamini\inst{10\orcidlink{0000-0003-1383-9414}}
   \and \\ G.~Di~Rosa\inst{9}\inst{,10}
   \and A.~Halkola\inst{11}
   \and A.~Mercurio\inst{12}\inst{,13}\inst{,14\orcidlink{0000-0001-9261-7849}}
   \and P.~Rosati\inst{9}\inst{,10\orcidlink{0000-0002-6813-0632}}
        }

   \institute{
   Finnish Centre for Astronomy with ESO (FINCA), Quantum, Vesilinnantie 5, University of Turku, FI-20014 Turku, Finland\\
   \email{stefan.schuldt@utu.fi}
   \and Department of Physics, P.O. Box 64, 00014 University of Helsinki, Finland
    \and Dipartimento di Fisica, Universit\`a  degli Studi di Milano, via Celoria 16, I-20133 Milano, Italy
    \and INAF -- IASF Milano, via A. Corti 12, I-20133 Milano, Italy
    \and Technical University of Munich, TUM School of Natural Sciences, Department of Physics,  James-Franck-Stra{\ss}e 1, 85748 Garching, Germany
    \and
    Max-Planck-Institut f{\"u}r Astrophysik, Karl-Schwarzschild Stra{\ss}e 1, 85748 Garching, Germany
    \and Instituto de F\'isica de Cantabria (CSIC-UC), Avda.~Los Castros s/n, 39005 Santander, Spain 
    \and INAF -- Istituto di Radioastronomia, via Gobetti 101, I-40129,
Bologna, Italy
    \and Dipartimento di Fisica e Scienze della Terra, Universit\`a degli Studi di Ferrara, Via Giuseppe Saragat 1, 44122 Ferrara, Italy
    \and INAF -- Osservatorio di Astrofisica e Scienza dello Spazio di Bologna, Via Piero Gobetti 93/3, 40129 Bologna, Italy
    \and Pyörrekuja 5 A, 04300 Tuusula, Finland 
    \and Università di Salerno, Dipartimento di Fisica ``E.R. Caianiello'', Via Giovanni Paolo II 132, I-84084 Fisciano (SA), Italy
    \and
    INAF -- Osservatorio Astronomico di Capodimonte, Via Moiariello 16, I-80131 Napoli, Italy
    \and
    INFN – Gruppo Collegato di Salerno - Sezione di Napoli,  Dipartimento di Fisica ``E.R. Caianiello'', Università di Salerno, via Giovanni Paolo II, 132 - I-84084 Fisciano (SA), Italy}

   \date{Received October 8, 2026}

 
  \abstract
   {We present the first inference of the Hubble constant and matter density parameter from a strongly lensed supernova by exploiting the mass model of a galaxy cluster enhanced by an extended surface brightness reconstruction. Building upon the recent mass model of the galaxy cluster MACS~J1149.5+2223, relying on the identified image positions only, we include the surface brightness distribution of the SN Refsdal host together with the measured time delays, to infer some cosmological parameter values within a flat $\Lambda$CDM framework. We consider two cosmological parameterizations, with either $H_0$ and $\Omega_\mathrm{m}$ or $H_0$ alone treated as free parameters, acknowledging the milder dependency on $\Omega_\mathrm{m}$ when incorporating additional constraints from the SN host. We explore four different mass-model realizations to assess some of the systematic uncertainties associated with the constraint weighting and source reconstruction. Although known as one of the main possible systematic effects, including influences from the external convergence, we deliberately keep the cluster total mass parametrization fixed to that of the previously published point-like mass model to enable a fair comparison. This reveals that incorporating 77,000 \HST pixels reduces the uncertainty on the inferred value of the Hubble constant by a factor of \f (when $\Omega_\text{m}$ is free), while all model realizations remain mutually consistent. Our final measurements are $H_0=$ \HoHoom~\kmsMpc and $\Omega_\text{m}=$ \omHoom, in agreement with previous analyses of SN Refsdal and with recent results from other strongly lensed supernovae. This corresponds to a substantially improved precision, reaching a remarkable \relerr uncertainty on the value of $H_0$ given our cluster total mass parameterization. The contributions of the uncertainties on the lens total mass and on the measured time delays to the final $H_0$ error are now comparable, indicating that future improvements in cluster-scale time-delay cosmography will increasingly depend on more precise time-delay measurements and continued control of systematic effects. In particular, with sub-percent time-delay values reachable at cluster-scale with time delays of years to decades, this mass model approach has the power to lead to unprecedentedly precise $H_0$ measurements. The methodology presented here is directly applicable to forthcoming strongly lensed supernovae discovered by the Rubin Observatory.}
   \keywords{gravitational lensing: strong -- methods: data analysis -- galaxies: clusters: general -- galaxies: clusters: individual: MACS J1149.5+2223 -- cosmological parameters}

   \maketitle

\section{Introduction}

Time-delay cosmography (TDC) is known as a robust, one-step method to measure the values of cosmological parameters, predominantly that of the Hubble constant, and is one of the key applications of strong gravitational lensing. It serves as a complementary approach to the SH0ES distance ladder \citep{riess22} and measurements from the Cosmic Microwave Background \citep[CMB;][]{planck20}, which currently yield significantly different results. This discrepancy, commonly referred to as the Hubble tension problem, remains one of the major open questions in modern cosmology \citep[e.g.,][]{verde19, divalentino21}. Originally proposed by \citet{refsdal64} with multiply imaged supernovae (SNe), TDC was first established with lensed quasars on both galaxy scale \citep[e.g.,][]{wong20, tdcosmo25} and cluster scale \citep[e.g.,][]{napier23, liu23}.

Finally, in 2014, the first strongly lensed supernova, now known as SN Refsdal, was discovered \citep{kelly15} behind the galaxy cluster MACS J1149.5+2223 (hereafter MACS1149). At that time, the SN was visible as a quadruply imaged source lensed by a single cluster member, while a fifth image was predicted by subsequent extensive modelling efforts \citep[e.g.,][]{grillo16, treu16} to appear about one year later, thanks to the lensing effect of the cluster as a whole. Thanks to extensive follow-up campaigns, including observations with the Hubble Space Telescope \citep[HST;][]{kelly15, lotz17}, the Grism Lens Amplified Survey from Space \citep[GLASS;][]{treu16}, and the Multi Unit Spectroscopic Explorer \citep[MUSE;][]{grillo16, schuldt24}, SN Refsdal was successfully exploited as a cosmological probe \citep[e.g.,][]{grillo18, kelly23, grillo24}. More recently, MACS1149 has been observed with the James Webb Space Telescope \citep[JWST, e.g.,][]{stiavelli23, morishita24, sarrouh26} to obtain additional photometric and spectroscopic data and remains an interesting target for establishing further TDC, studying the host galaxy at $z=1.49$ and investigating other background sources, such as JD1 at $z=9.1$ \citep{zheng12, Hashimoto18}.

Besides SN Refsdal, several other strongly lensed SNe have been discovered. So far, three additional systems, SN Encore \citep{pierel24}, a sibling of SN Requiem, SN H0pe \citep{frye24}, and SN Winny \citep{johansson25, taubenberger26} have led to measurements of $H_0$ \citep{pascale25, bazzanini26,  suyu26, pierel26, moertsell26, townsend26, johansson26_winny}. Notably, a counter-image of SN Requiem is expected to appear, depending on the actual value of $H_0$, in 2026 or 2027 \citep{suyu26}, turning this system into an extremely powerful cosmological probe. While most useful systems are so far on cluster scale, with the advent of the Legacy Survey of Space and Time (LSST) at the Rubin Observatory \citep{ivezic08}, combined with high-resolution optical and near-infrared imaging from Euclid \citep{laureijs11}, Roman \citep{green12, perkins24_roman}, and the Chinese Space Station Telescope \citep{gong19}, an increasing number of strongly lensed SNe are expected to be discovered \citep[e.g.,][]{goldstein19}.


However, despite these newer candidates, SN Refsdal remains the best-studied strongly lensed SN system and the ideal test case for establishing new modeling codes and analysis procedures. Therefore, \citet{schuldt24} presented a new mass model incorporating additional VLT/MUSE observations in the northern part of the galaxy cluster. This extended the catalog of known cluster members brighter than 24 mag (AB) in the HST F160W band to 308 systems and increased the set of strong-lensing constraints to 63 multiple images belonging to 18 knots of the SN Refsdal host galaxy ($z_\text{SN} = 1.489$), together with 43 multiple images from 16 other strongly lensed sources spanning the redshift range $1.24 \leq z \leq 5.98$. While this model is based solely on the identified image positions (a so-called ``point-like'' mass model), \citet{schuldt26} enhanced it by incorporating the source surface brightness distribution of 77,000 HST pixels, similarly to the pioneering work by \citet{acebron24}, who modeled the strongly lensed quasar host in the galaxy cluster SDSS J1029+2623. 
This approach allows the extended surface brightness information of the lensed sources to provide additional constraints on the gravitational potential beyond discrete image positions \citep[e.g.,][]{warren03, vegetti09a, suyu06a}. As a result, it led to a significant reduction in the statistical uncertainty, making the time-delay uncertainty a comparable source of error while also highlighting the need to account for systematic uncertainties. While \citet{schuldt26} presented an enhanced mass model of MACS1149, together with the expected improvements in the mass-model parameters, inferred magnifications, and reconstruction of the cluster total mass distribution, this work presents the resulting cosmological parameter measurements. Specifically, we consider the flat Lambda Cold Dark Matter ($\Lambda$CDM) model in two cases: first with the matter density parameter fixed to $\Omega_\text{m}=0.3$, leaving $H_0$ as the only free cosmological parameter, and second with both $H_0$ and $\Omega_\text{m}$ treated as free parameters. This work therefore represents the first cosmological analysis based on a surface-brightness-informed mass model of a galaxy cluster, providing an assessment of the achievable precision from current strong-lensing constraints and the remaining limitations from time-delay and some systematic uncertainties.

The structure of the paper is as follows. We summarize the mass model setup and discuss modifications compared to the model presented by \citet{schuldt26} in Sect.~\ref{sec:model}. Section~\ref{sec:inference} details the cosmological inference procedure and which model variations we considered. In Sect.~\ref{sec:results}, we then describe the results, before concluding in Sect.~\ref{sec:conclusion}.

\FloatBarrier
\section{Lens mass model}
\label{sec:model}

We first provide a short summary of the adopted mass parameterization and our different schemes how to weight the multiple image positions and the source surface brightness distribution, followed by a discussion of the modifications relative to \citet{schuldt26}. For further details, we refer to \citet{schuldt24, schuldt26} and to Sect.~\ref{sec:inference} for the incorporation of the time delays and the procedure used to perform the cosmological inference. The presented lens mass model is constructed using the \GLEE software \citep{suyu10a_GLEE, suyu12b_GLEE}, complemented by \gleetoolspy \citep{schuldt23b}, which provides additional analysis and post-processing routines.

\subsection{Mass parametrization and weighting schemes}
\label{sec:model:parametrization}

The adopted mass parameterization follows directly that of \citet{schuldt24} to allow for a fair and direct comparison between the previously presented mass model constructed using the state-of-the-art procedure and this work, which additionally incorporates the source surface brightness distribution. For this reason, we do not include here a few recently identified multiple images or new redshift measurements, for instance from \citet{sarrouh26}.

In short, the cluster total mass distribution is parameterized as follows. The overall cluster dark-matter distribution is described by three pseudo-isothermal elliptical mass profiles \citep{kassiola93}, two of which are centered in the cluster core. An additional fourth halo, assumed to be circular, is included to represent the galaxy group halo located in the northern part of the cluster. Additionally, we include 308 securely identified cluster members, using dual pseudo-isothermal elliptical (dPIE) mass profiles \citep{eliasdottir07, suyu10a_GLEE} with vanishing core radii. While 306 cluster members are assumed to be spherical and their Einstein and truncation radii are scaled according to their luminosity measured in the \HST F160W band, two members are modeled as elliptical (i.e., with free axis ratios and position angles) and excluded from the scaling relations (i.e., with free Einstein and truncation radii) to increase the model flexibility near the SN images. This results in a total of 31 free mass parameters, in addition to three free redshifts of secure multiple images without spectroscopic redshifts at the time when the initial mass model was built. These 34 parameters, together with the cosmological parameters (see Sect.~\ref{sec:inference}), are optimized through simulated annealing \citep{suyu10a_GLEE, suyu12b_GLEE} and subsequently sampled using emcee \citep{foreman-mackey13}.

As in \citet{schuldt26}, these free parameters are constrained by a combination of multiple image positions and the source surface brightness distribution of the SN host galaxy. We consider the four different combinations from \citet[referred to as \Ew, \Sw, \Ewo, and \Swo]{schuldt26} to investigate the impact of weighting schemes on the inferred cosmological parameters. All four model variations include the five SN Refsdal image positions and 43 multiple image positions from 16 background sources at redshifts different from that of the SN. The additionally identified 63 knots of the SN host are incorporated in the models \Ew and \Sw, while they are excluded in \Ewo and \Swo to test their impact. Furthermore, while all four versions include the same 77,000 \HST pixel intensities \citep[see Fig.~1 in][]{schuldt26}, we test two weighting schemes for their contribution to the $\chi^2$. In models \Sw and \Swo, we scale the image position uncertainties and error map such that their contribution to the total $\chi^2$ is comparable, whereas for \Ew and \Ewo we retain the image position uncertainties from \citet{schuldt24}, resulting in the extended image dominating the optimization. As in \citet{schuldt26}, we use the ``point-like'' mass model from \citet{schuldt24} as a reference model for a direct comparison, which relies solely on the identified image position coordinates.

\subsection{Adopted point spread function}
\label{sec:model:PSF}

\citet{schuldt26} adopted a simplified point spread function (PSF) to speed up the convolution. However, as already noted in \citet{schuldt26}, an inference of physical quantities requires an empirically motivated PSF. Stars, as point sources, provide a direct probe of the PSF shape, and combining multiple stars helps mitigate local variations and diffraction spikes. Consequently, if a PSF is not provided by the imaging survey, it is typically approximated by stacking nearby stars observed within the same dataset as the science target.

However, in our case, we use a difference image $I_\text{dif}$ constructed from two HST filters to subtract the cluster light by scaling one band, making the selection of an appropriate PSF non-trivial. In detail, we scale the F606W image by a constant $\alpha$ to match the cluster member intensities in the F435W band \citep[see details in][]{schuldt26}, such that the difference image contains predominantly the flux from the SN host in the considered regions. To obtain the PSF used in this analysis, we define
\be
I_k = \text{PSF}_k \circledast L_k
\ee
as the observed intensity in filter $k$, expressed as the convolution between the PSF and the intrinsic (unconvolved) intensity $L$ of the given filter $k$. Since we construct the difference image between the F435W and F606W filters, we have
\bea
I_\text{dif} &= &I_\text{F435W} - \alpha I_\text{F606W} \nonumber \\
             &= &\text{PSF}_\text{F435W} \circledast L_\text{F435W} - \alpha \text{PSF}_\text{F606W} \circledast L_\text{F606W} \nonumber \\
             &\sim & \text{PSF}_\text{F435W} \circledast (L_\text{F435W} - \alpha L_\text{F606W} )\\
             &=& \text{PSF}_\text{F435W} \circledast L_\text{dif} \nonumber \, .
\eea

Here we assume that the PSFs of the two filters are sufficiently similar. This assumption is supported by the following arguments: (1) $\alpha$ is very small, such that the difference image contains predominantly flux from the F435W band; (2) the two filters partly cover the same wavelength range, resulting in comparable PSFs, although small differences in the PSF full width at half maximum (FWHM) have been reported by \citet{queirolo25} \citep[noting that neither F435W nor F606W were considered in][]{queirolo25}; (3) both filters cover the optical wavelength range, where the PSF variation is expected to be smaller than in the near-infrared; and (4) the PSF is approximated from three stars following the procedure of \citet{ertl23}, such that the derived PSF remains an approximation.

As a final test of the PSF impact, we ran short test chains using three different variations. First, we adopted the approximate PSF as in \citet{schuldt26}, we updated the error map (see Sect.~\ref{sec:model:errormap}), or updated the error map and the PSF to the stacked stars from the F435W. After several short iterative sampling chains, we selected 100 random iterations of the last chain and computed the predicted time delays. Based on this test, the PSF change has only a minor impact on the time delays. The relatively small sensitivity to the PSF compared to galaxy-scale systems may be related to the additional constraints provided by the multiple image positions at different redshifts. Therefore, for the models presented here, we use a PSF constructed from three stacked stars in the F435W band.

\subsection{Error map}
\label{sec:model:errormap}

Since we model the difference image of two HST bands as described in Sect.~\ref{sec:model:PSF}, we first construct the error maps for each band individually using \gleetoolspy, following the procedure described in \citet{schuldt26}. These are then combined as
\be
\sigma_{\text{tot,}j} = \sqrt{\sigma_{\text{tot, F435W,}j}^2 + \alpha^2 \sigma_{\text{tot, F606W,}j}^2} \, ,
\label{eq:errmap}
\ee
thereby taking the scaling factor $\alpha$ into account.

\subsection{Mitigating cluster member light remnants}

\begin{figure}[t!]
    \centering
    \includegraphics[width=\linewidth]{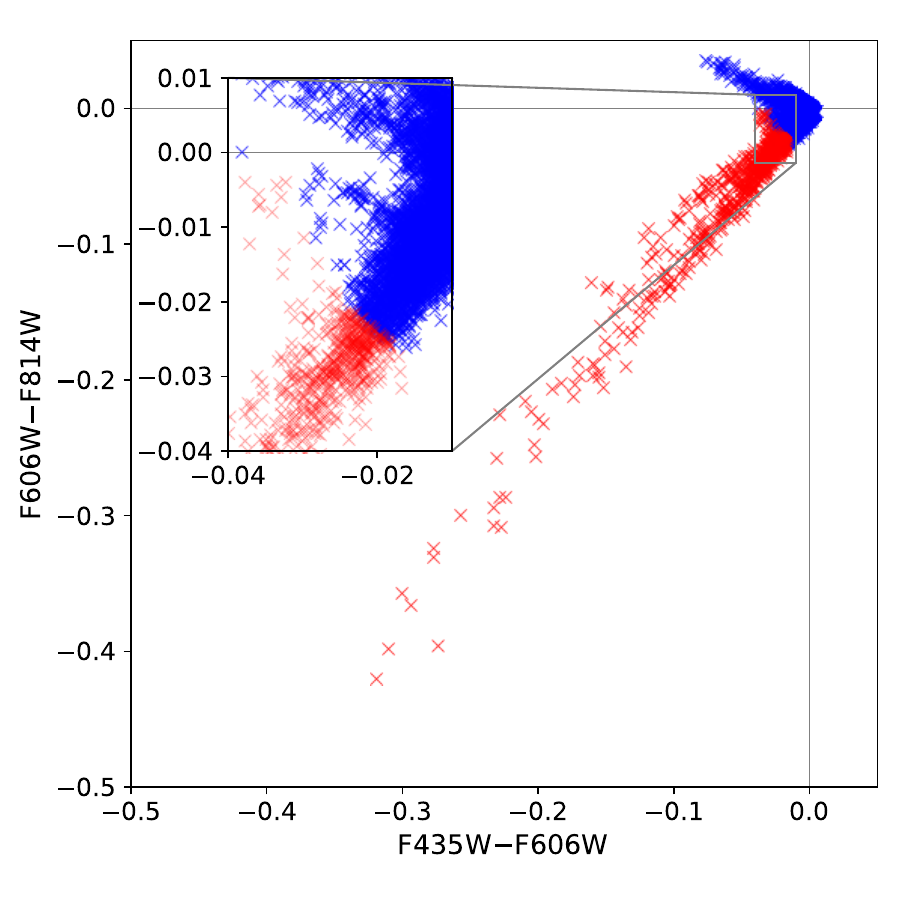} \\\includegraphics[width=\linewidth, trim = 50 0 50 0, clip]{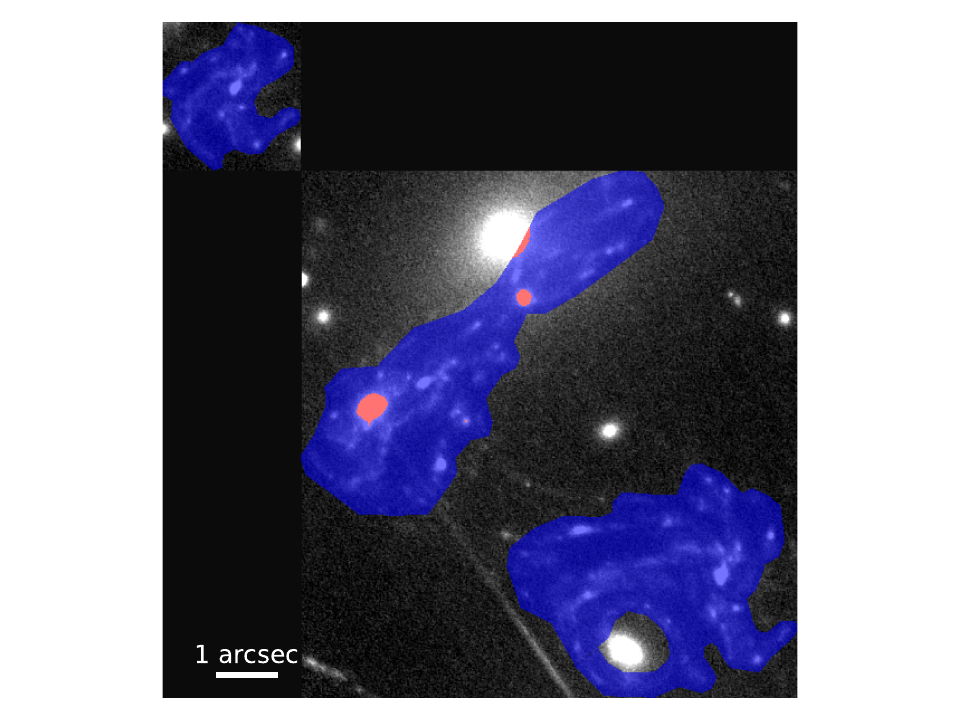}
    \caption{\textit{Top:} Color-color plot of all 77,000 HST pixels used for the source surface brightness reconstruction. Red crosses represent the 798 pixels slightly contaminated by the light from some cluster members ($\sim1\%$), while the color positions of the other pixels are shown in blue. To mitigate the remaining cluster light contamination, the corresponding uncertainties are now boosted by a factor of 10. \textit{Bottom:} Spatial distribution of the pixels containing some residual cluster light following the same color coding as above with the HST F606W image in the background.}
    \label{fig:colorcolor_test}
\end{figure}

To further mitigate residual contamination from cluster members, we created a color-color diagram of the considered HST pixels (F435W$-$F606W versus F606W$-$F814W), as shown in Fig.~\ref{fig:colorcolor_test} (top panel). Although the bulge of the SN host galaxy appears redder than the spiral arms, these pixels occupy a clearly distinct region in color-color space from pixels contaminated by cluster-member light. In particular, while pixels belonging to the host bulge have positive values in the F606W$-$F814W color, pixels affected by cluster light fall predominantly into the negative range. This allows us to clearly separate the two populations. We define pixels affected by cluster-member light as those with (F606W$-$F814W) $\leq 0$ and a distance larger than 0.03 in color-color space from the median of all pixels, where the distance threshold was determined empirically based on our tests. This corresponds to 798 pixels, or approximately 1\% of the 77,000 HST pixels which are shown in red in Fig.~\ref{fig:colorcolor_test}. This small fraction further demonstrates the effectiveness of the cluster-light subtraction. Although excluding these contaminated pixels affects the total $\chi^2$ only by the expected amount, we follow the approach of \citet{acebron24} and instead increase their uncertainties by a factor of 10 to mitigate their influence.


We further increase the uncertainties of pixels containing the SN images because the source reconstruction grid cannot adequately represent unresolved point sources such as the SN images and because the SN image positions, together with their measured time delays, are already included as independent constraints in all considered mass models (see Sect.~\ref{sec:inference}). This avoids counting the same information twice, once through the pixel intensities and once through the image positions and time delays, although some degree of correlation remains for the \Sw and \Ew models. Nevertheless, we retain these models to assess the associated systematic uncertainties.

\subsection{Test of a color gradient}

Since we rely on a difference image, we tested whether the image subtraction introduces a color gradient. For this purpose, we selected 10 cluster members located near the region of interest (i.e. around the SN host) and measured the integrated flux within a circular aperture of 1\arcsec diameter in both the F435W and F606W bands, taking the scaling factor $\alpha$ into account. Furthermore, we selected 10 random positions devoid of detectable astrophysical sources and performed the same test. Comparing these 20 measurements between the two filters revealed the expected scatter, but no systematic trend indicative of an introduced color gradient. We therefore conclude that the difference image is well suited for the mass modeling.

\FloatBarrier
\section{Cosmological inference}
\label{sec:inference}

In this work, we consider two different cosmological models: one with a free Hubble constant, $H_0$, while fixing $\Omega_\text{m} \equiv 0.3$ (hereafter referred to as the ``H0'' models), and one in which both $H_0$ and $\Omega_\text{m}$ are free to vary (labeled ``H0+om''), resulting in a total of eight different model realizations. In all cases, we assume a flat $\Lambda$CDM cosmology, i.e., $\Omega_\text{m}=1-\Omega_\text{DE}$, with the dark-energy equation-of-state parameter fixed to $w=-1$. Uniform priors are adopted over the ranges $20$--$120\,\kmsMpc$ and $0$--$1$ for $H_0$ and $\Omega_\text{m}$, respectively.

To constrain these parameters, we incorporate the measured time delays into the mass models using the ``combined'' measurements of \citet{kelly23_timde_delays} with symmetrical errors, namely $\Delta t_\text{S2:S1}=(9.9\pm4.0)$ days, $\Delta t_\text{S3:S1}=(9.0\pm4.2)$ days, $\Delta t_\text{S4:S1}=(20.3\pm6.4)$ days, and $\Delta t_\text{SX:S1}=(376.0\pm5.6)$ days. Here $\Delta t_{i:\text{S1}}$ specifies the time delay between image $i$ and S1, following the naming convention of \citet{kelly23_timde_delays}. We adopt the quoted $1\sigma$ uncertainties directly, such that the contribution of the time delays to the total $\chi^2$ remains comparatively small relative to that of the extended surface brightness and, particularly for the \Sw and \Swo models, the multiple image positions. The unequal contributions of the different observables to the total $\chi^2$ are a well-known aspect of TDC and one of the motivations for exploring different weighting schemes.

The optimization of the mass models including the time delays and cosmological parameters is performed with \GLEE \citep{suyu10a_GLEE, suyu12b_GLEE}. Following \citet{schuldt26}, we first optimize each model using a source grid of $100\times100$ pixels. Given the comparable small contribution from the time delays to the total $\chi^2$, we also ran iteratively chains by fixing either some or all of the mass model parameters and free redshifts to better optimize the cosmological parameters depending on the time delays. After the Markov chains had stabilized for all eight model realizations, we ran ten final chains for each model, adopting different random seeds and varying the source-grid size between $95\times95$ and $105\times105$ pixels to account for possible systematic effects associated with the source grid size. Here we varied again all 35 or 36 parameters for the H0 and H0+om models, respectively, while we confirmed that a fixed mass parametrization results in consistent cosmological results. In detail, for each model realization, we first performed ten short chains of 4,000 iterations (from 80 walkers with 50 samples), which were combined to estimate the final covariance matrix. Using this covariance matrix, we then ran ten chains of 40,000 iterations each (from 80 walkers and 500 samples), resulting in a stacked chain of 400,000 samples per model. 

To assess convergence, we computed the rank-normalized and folded split-$\hat{R}$ diagnostic for each model parameter across the ten independent chains, following \citet{vehtari21_rhat}. For all model variations, the resulting $\hat{R}$ values did not exceed 1.025 and the mean of the $\hat{R}$ values of all 35 or 36 parameters never exceeded 1.015. This indicates good convergence of the parameters relevant for the cosmological inference \citep{vehtari21_rhat}.
The finally inferred cosmological parameter values from these eight combined chains are reported in Table~\ref{tab:cosmo}.

\begin{table}[]
    \centering
    \caption{Cosmological values from the final chains. }
    \begin{tabular}{l|cc}
         Model & $H_0 [\kmsMpc]$ & $\Omega_\text{m}$\\ \hline \hline
\S24 (H0)    &  $65.4_{-3.7} ^{+ 3.7}$ &  -- \\ 
\S24 (H0+om) &  $65.4_{-3.7} ^{+ 3.8}$  &  $0.34_{-0.15} ^{+ 0.18}$  \\ \hline 
\Ew (H0)    &  $65.8_{-2.0} ^{+ 1.9}$ &  -- \\ 
\Ew (H0+om) &  $66.6_{-1.8} ^{+ 1.9}$  &  $0.31_{-0.10} ^{+ 0.09}$  \\ 
\Ewo (H0)    &  $65.9_{-1.9} ^{+ 1.8}$ &  -- \\ 
\Ewo (H0+om) &  $65.4_{-1.8} ^{+ 1.8}$  &  $0.30_{-0.10} ^{+ 0.08}$  \\ 
\Sw (H0)    &  $66.3_{-1.8} ^{+ 1.9}$ &  -- \\ 
\Sw (H0+om) &  $65.2_{-1.8} ^{+ 1.8}$  &  $0.28_{-0.09} ^{+ 0.08}$  \\ 
\Swo (H0)    &  $66.1_{-1.7} ^{+ 1.5}$ &  -- \\ 
\Swo (H0+om) &  $64.7_{-1.5} ^{+ 1.7}$  &  $0.30_{-0.09} ^{+ 0.08}$  \\ \hline
Final (H0)    & $ 66.0 ^{+ 1.6 }_{ -1.8 }$ & -- \\ 
Final (H0+om) & $ 65.0 ^{+ 2.0 }_{ -1.8 }$ & $ 0.3 ^{+ 0.08 }_{ -0.09 }$
    \end{tabular}
    \label{tab:cosmo}
\end{table}

Since the \Sw and \Ew models, as well as the \Swo and \Ewo models, are based on identical data sets and differ only in their weighting schemes, we additionally combine each pair with the same free cosmological parameters into a single stacked chain. As this work presents the first cosmological inference based on an extended source reconstruction for a cluster-scale lens, the adopted weighting of the different $\chi^2$ contributions constitutes an additional and unexplored source of systematic uncertainty. Furthermore, the \Sw and \Ew models intentionally double count the information from the SN host by including it both through the pixel intensities and through the identified multiple image positions. We therefore treat these models as a systematic test. Consequently, we adopt the combined \Ewo+\Swo chain as our fiducial result, while using the \Ew+\Sw combination to estimate the associated systematic uncertainty. For a parameter $p\in \{H_0,\Omega_\text{m} \}$, we define

\begin{equation}
\sigma_{\rm sys}
=
\left|p_\text{Ew+Sw}-p_\text{Ewo+Swo}\right|,
\end{equation}
as the systematic uncertainty raised by the adopted weighting scheme. Although we assumed that the true value lies approximately midway between the two weighting schemes, we estimate the systematic uncertainty as their direct difference to be conservative. We then compute the total uncertainty as
\begin{equation}
\sigma_{{\rm tot},j} = \sqrt{
\sigma_{{\rm stat},j}^2
+
\sigma_{{\rm sys}}^2
}~,
\end{equation}
where $j\in \{+,-\}$ denotes the upper and lower statistical uncertainties obtained from the 84th and 16th percentiles of the stacked chain, respectively. The final values reported in Table~\ref{tab:cosmo} are therefore

\begin{equation}
p^{+\sigma_{\rm tot,+}}_{-\sigma_{\rm tot,-}}.
\end{equation}

For comparison, we also report the inferred cosmological parameter values obtained from the \S24 point-like mass model in Table~\ref{tab:cosmo}. Here, the parameters are sampled with \GLEE using Markov chain Monte Carlo and optimized through simulated annealing. We ran ten independent chains with different random seeds, each consisting of 500,000 iterations, and combined them into a single stacked chain. We emphasize that the quoted uncertainties should be interpreted as statistical uncertainties only, despite the use of multiple random seeds.

\FloatBarrier
\section{Results and Discussion}
\label{sec:results}

\begin{figure}[t!]
    \centering
    \includegraphics[width=\linewidth]{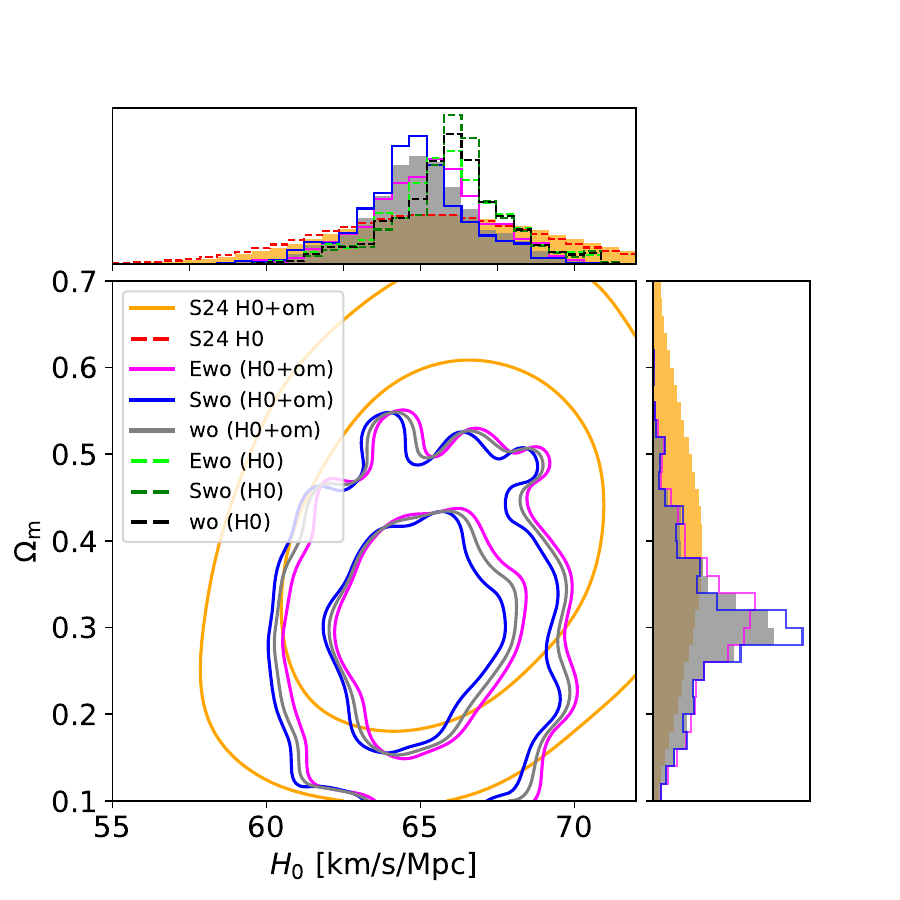}\\\includegraphics[width=\linewidth]{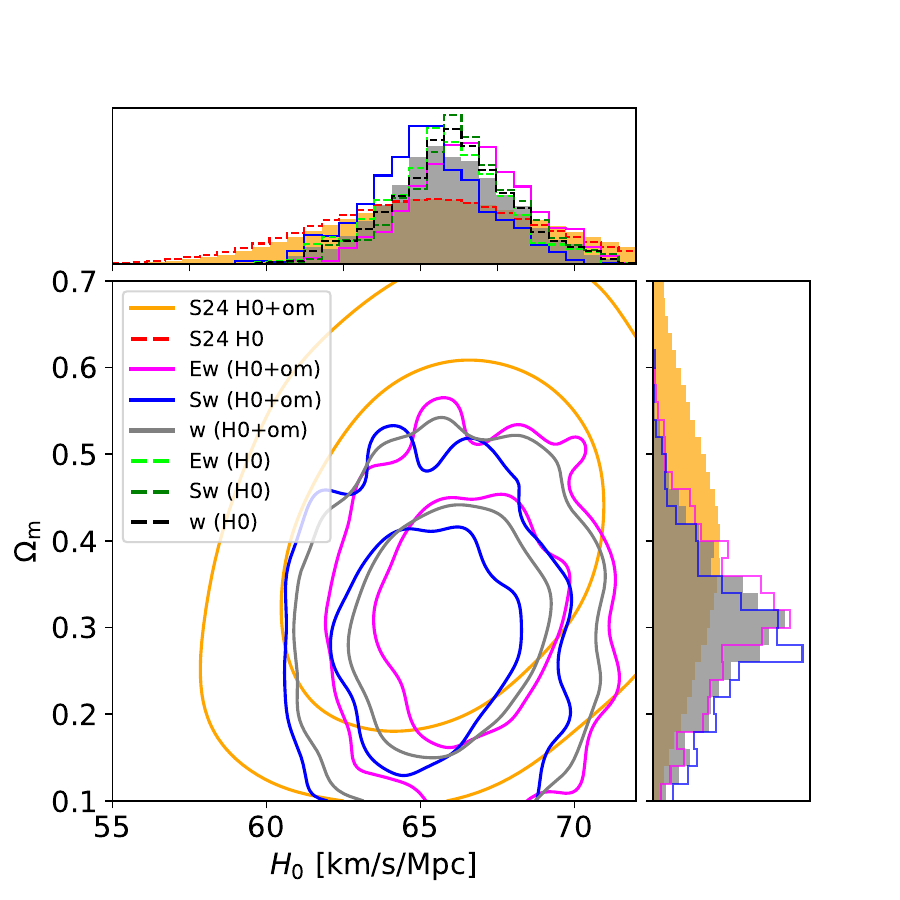}
    \caption{2D contour plots (1$\sigma$ and 2$\sigma$) and normalized histograms of the two inferred cosmological parameters $H_0$ and $\Omega_\text{m}$. The top panel shows the results from the point-like mass models \S24 and from those excluding the knots of the SN host, while the bottom panel shows those from \S24 and from those with the SN host knots. Here w and wo are abbreviations for \Ew+\Sw and \Ewo+\Swo, respectively. The impact from incorporating the source surface brightness distribution is clearly evident.}
    \label{fig:H0_om}
\end{figure}

\begin{figure}
    \centering
    \includegraphics[width=\linewidth]{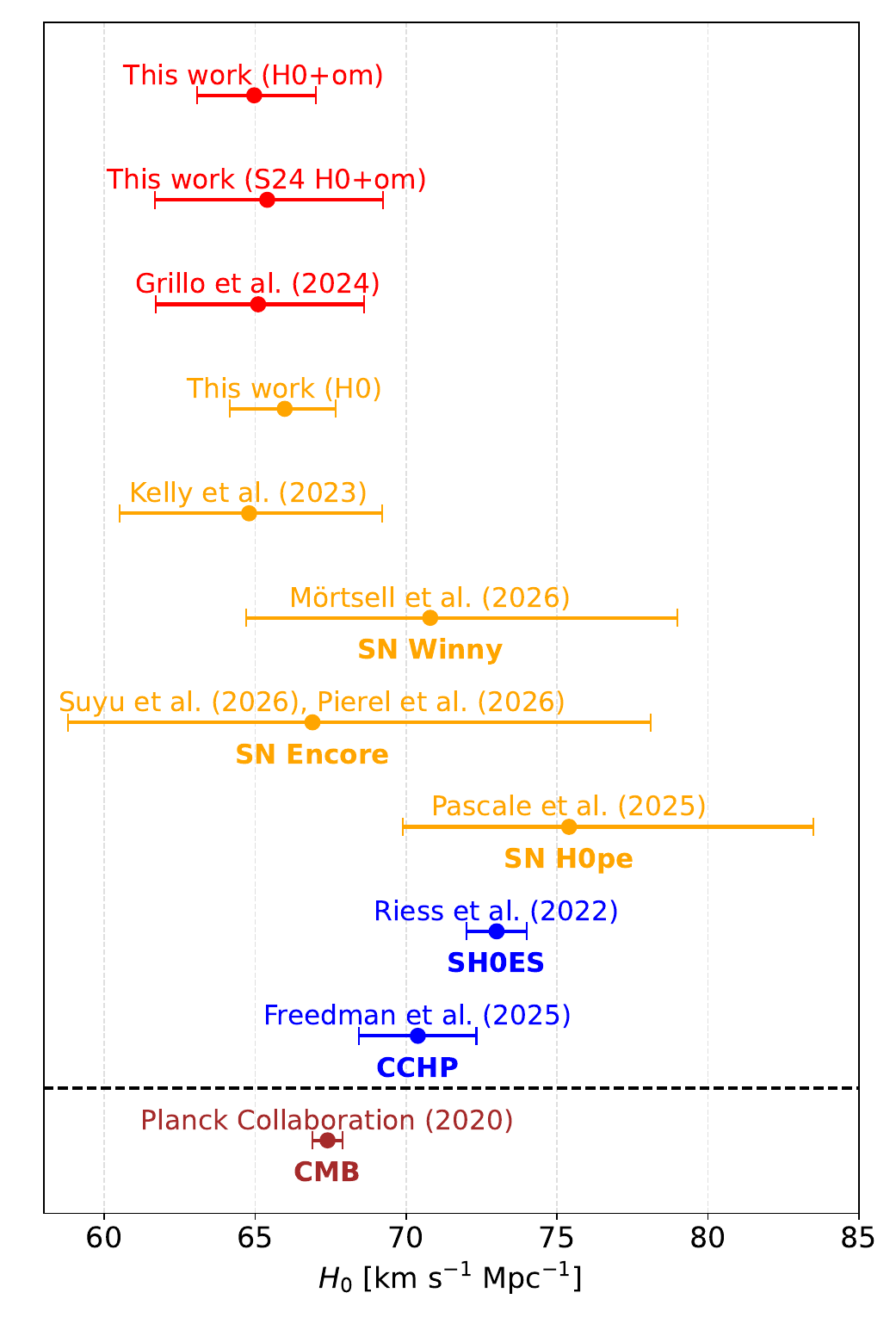}
    \caption{Measured values of the Hubble constant from this work compared to the literature. Results from lensed SNe with only varying $H_0$ are shown in orange, while those with more flexible cosmological model are in red. Our results from SN Refsdal are well in agreement with previous results on the same system \citep{kelly23, grillo24}, as well as the values inferred from the CMB \citep[brown,][]{planck20}, the Chicago-Carnegie Hubble Program \citep[CCHP]{freedman25}, and the SN Encore \citep{pierel26, suyu26} and SN Winny systems \citep{moertsell26}.}
    \label{fig:H0_probes}
\end{figure}

The obtained cosmological parameter values from the different mass model variations used to assess systematic uncertainties, together with our final combined measurements, are presented in Table~\ref{tab:cosmo} and visualized in Fig.~\ref{fig:H0_om} as contour plots and 1D histograms for the two considered cosmological parameters. Figure~\ref{fig:H0_probes} places our final $H_0$ measurement in the context of previous measurements from SN Refsdal \citep{kelly23, grillo24} and other strongly lensed SNe \citep{pascale25, suyu26, pierel26, moertsell26}, as well as complementary probes based on the CMB \citep{planck20} and the cosmic distance ladder \citep[SH0ES,][]{riess22}.

Most notably, all of our model variations yield mutually consistent results within $1\sigma$, and are likewise consistent with previous measurements based on SN Refsdal. This already indicates that the different weighting schemes introduce only a limited level of systematic uncertainty. Our measurements are also consistent with those obtained from SN Encore by \citet{pierel26, suyu26} and with the CMB determination from \citet{planck20}. However, owing to the improved precision achieved by directly incorporating the surface brightness distribution of the SN host galaxy, the resulting measurements deviate more from those inferred from SN H0pe \citep{pascale25} and from the SH0ES distance ladder \citep{riess21}.

For the point-like mass model (\S24), we obtain a statistical uncertainty of 5.8\% on $H_0$, consistent with the results of \citet[6.5\%]{kelly23} and \citet[5.3\%]{grillo24}. While the mass model presented by \citet{schuldt24} and adopted for the \S24 cosmological analysis is based on the model of \citet{grillo16} used by \citet{grillo24}, it significantly extends the set of multiple-image constraints and, in particular, incorporates the galaxy group identified in the northern part of the cluster. As a result, the global cluster mass distribution is less sensitive to the knots of the SN host galaxy, making the slightly larger uncertainty on the inferred cosmological parameters plausible. We further note that the inferred values of $\Omega_\text{m}$ are also in good agreement with those reported by \citet{grillo24}, while we find a low sensitivity to $\Omega_\mathrm{m}$ possibly resulting in an underestimation of the uncertainty. This low dependency can be well explained by the fact that $\Omega_\mathrm{m}$ is independent of the SN host surface brightness dominating the mass model optimization and only dependent on the distance ratios from multiple images at different redshifts.

As evident from Figs.~\ref{fig:H0_om} and \ref{fig:H0_probes}, incorporating the extended surface brightness distribution of the SN host substantially improves the precision of the cosmological inference. For our final measurement, we obtain a relative uncertainty of \relerr, corresponding to an improvement by a factor of \f compared to the point-like mass model (\S24). Also the uncertainty on $\Omega_\mathrm{m}$ appears to be significantly reduced, but likely underestimated, given the weaker dependency on the multiple image positions once the extended surface brightness of the SN host is included. We emphasize that this uncertainty already includes some systematic contributions, namely those arising from different weighting schemes and source-grid sizes. We deliberately retain the cluster total mass parametrization of \citet{schuldt24}, to enable a fair and direct comparison, although this implies that we do not account for additional sources of systematic uncertainty related to the specific mass modeling details. Furthermore, the mass-sheet degeneracy (MSD) is partly mitigated by the inclusion of additional source redshifts, which provide complementary lensing constraints, although its impact is limited given the additional constraints from the SN host surface brightness. We therefore acknowledge the residual MSD as an additional source of systematic uncertainty that is not further accounted for here. Moreover, the consistency among the individual model realizations demonstrates that these methodological choices do not significantly affect the inferred cosmological parameters.

\begin{figure*}
    \centering
    \includegraphics[width=\linewidth]{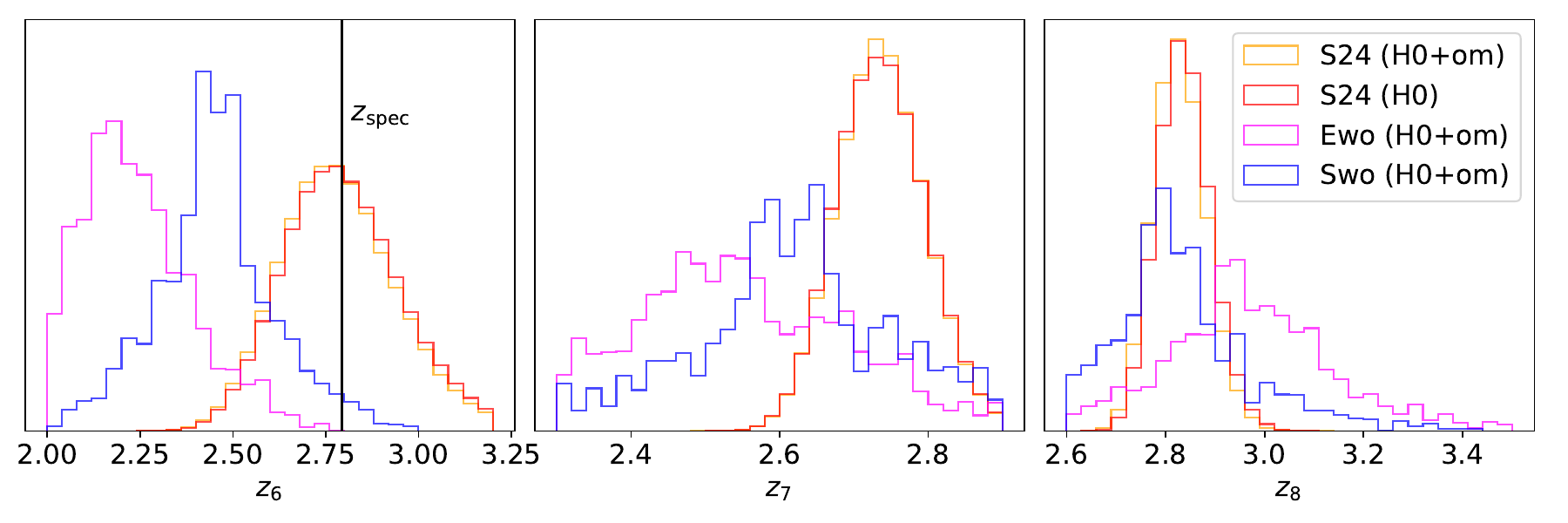}
    \caption{Redshift distribution for families 6, 7 and 8 predicted by the models \S24 (H0+om), \S24 (H0), \Ewo (Ho+om) and \Swo (H0+om) as representative examples. The redshift of family 6 is now confirmed by JWST to be at $z_\text{spec}=2.7927$ \citep{sarrouh26}, as indicated by the black line.}
    \label{fig:z}
\end{figure*}

Although this work focuses on the inferred cosmological parameters rather than on the recovered mass model parameters, we note that the latter are overall broadly consistent with those presented by \citet{schuldt24}. Furthermore, as discussed in Sect.~\ref{sec:model:parametrization}, we intentionally did not incorporate the recently measured redshift of family 6, $z=2.7927$, reported by \citet{sarrouh26}. Figure~\ref{fig:z} shows the posterior distributions of the three source families with free redshifts for four representative mass models. As already noted by \citet{schuldt26}, model \Swo, which combines multiple-image positions and extended source surface brightness with comparable weights, yields considerably narrower redshift distributions than the point-like model \S24 and the extended-image-dominated model \Ewo. This is expected because the uncertainties assigned to the multiple-image positions, including those of family 6, are effectively reduced in the \Swo weighting scheme. While the median value of $z_6$ inferred from the \S24 model agrees well with the spectroscopic JWST measurement, the posterior distributions from the \Swo and \Ewo models are shifted toward lower values, particularly for family 6. Given its weighting scheme focusing on the extended image and downweighting the contribution of the multiple image positions, the \Ewo model is not well suited for constraining source redshifts, as the optimization is effectively driven by the reconstruction of the SN host galaxy. This illustrates that a smaller statistical uncertainty, as reached with e.g., model \Ewo compared to \S24, does not necessarily imply a more accurate redshift prediction when the optimization is dominated by a subset of the constraints. It highlights once more the necessity now to incorporate also systematic uncertainties.

\section{Conclusions}
\label{sec:conclusion}

We inferred, for the first time, the values of some cosmological parameters from a strongly lensed supernova using the source surface brightness reconstruction of the SN host galaxy. We considered two cosmological models: one varying only the Hubble constant (H0 models) and one assuming a flat $\Lambda$CDM cosmology with both $H_0$ and $\Omega_\text{m}$ treated as free parameters (H0+om models). For each cosmological model, we explored four different mass model realizations to assess some systematic uncertainties. Specifically, we adopted two different weighting schemes, either assigning dominant weight to the SN host pixels or balancing the contributions from the identified image positions and the \HST\ pixels. This was achieved by scaling the uncertainties associated with the multiple image positions and the error map of the SN host by constant factors. Furthermore, we considered models that either included or excluded the identified SN host knots as additional constraints. This allowed us to estimate the systematic uncertainties related to the specific weighting choice, which are incorporated into the final reported values, and, by additionally performing the cosmological inference with the point-like mass model, to provide a fair and direct comparison with the current state of the art. All models included the SN image positions together with the measured time delays from \citet{kelly23_timde_delays}, even when the SN host knots were excluded, and adopted the mass parameterization of \citet{schuldt24}.

The final inferred values are $H_0=\HoHoom~\kmsMpc$ and $\Omega_\text{m}=\omHoom$, and $H_0=\Ho~\kmsMpc$ when fixing $\Omega_\text{m}\equiv0.3$, with the corresponding posterior distributions shown in Fig.~\ref{fig:H0_om}. Incorporating the 77,000 \HST pixels through the source surface brightness reconstruction substantially improves the cosmological precision, reducing the final uncertainty on $H_0$ by a factor of approximately \f compared to the point-like mass model (\S24). This demonstrates clearly that strong lensing mass models should exploit the source surface brightness reconstruction, even on cluster scales.

Since the mass parameterization and the set of multiple-image constraints are otherwise identical, this improvement can be attributed directly to the inclusion of the extended surface brightness information. Furthermore, as the uncertainty from the mass model is now comparable to that of the longest measured time delay, $\Delta t_\mathrm{SX:S1}$, we conclude that the time-delay measurements have become a dominant contribution to the overall uncertainty budget. Next, systematic uncertainties associated to the total mass modeling details, not always fully explored for known strongly lensed SNe, will have to be properly quantified \citep[as done by][]{grillo20}. Furthermore, given the increasing precision on measurements from cluster-scale systems and notable difference from galaxy-scale ones, the overall external convergence will have to be considered carefully in the models. If the lines of sight of galaxy clusters were under-dense, compared to a typical line of sight, this would result in underestimates of the value of $H_0$ from cluster lenses. Both aspects need to be addressed with future measurements and models, but they are beyond the scope of this work.   

Thanks to the increased number of strongly lensed SN discoveries over the past few years, this methodology is expected to play a significant role in high-precision cosmography on cluster scale, similarly as it does on galaxy scale. In particular, the shown improvement is very timely in view of the imminently expected reappearance of SN Requiem, after approximately ten years from its first detection \citep{rodney21, suyu26}. This will likely provide an exceptionally precise time-delay measurement. Combined with mass models incorporating source surface brightness reconstruction, as presented here, this system has the potential to enable a few-percent measurement of the value of the Hubble constant. 

Finally, the forthcoming high-resolution imaging from Euclid, Roman, and the Chinese Space Station Telescope will substantially expand the available optical and near-infrared data sets, enabling the discovery and characterization of an unprecedented number of strong gravitational lenses. Combined with the transient discoveries from LSST at the Rubin Observatory, these facilities are expected to open a new generation of strongly lensed transient discoveries and increasingly precise time-delay cosmography.

\begin{acknowledgements}
SS has received funding from the European Union’s Horizon 2022 research and innovation programme under the Marie Skłodowska-Curie grant agreement No 101105167 — FASTIDIoUS. PB acknowledges financial support from the Italian Space Agency (ASI) through contract ``Euclid - Phase E'', and INAF Grants ``The Big-Data era of cluster lensing'' and ``Probing Dark Matter and Galaxy Formation in Galaxy Clusters through Strong Gravitational Lensing''.
This research is based on observations made with the NASA/ESA \textit{Hubble} Space Telescope obtained from the Space Telescope Science Institute, which is operated by the Association of Universities for Research in Astronomy, Inc., under NASA contract NAS 5-26555. Based on observations collected at the European Southern Observatory under ESO programmes 294.A-5032 and 105.20P5.001.

Software Citations:
This work uses the following software packages:
\href{https://github.com/astropy/astropy}{\texttt{Astropy}}
\citep{astropy1, astropy2},
\href{https://github.com/dfm/emcee}{\texttt{Emcee}}
\citep{foreman-mackey13},
\href{https://github.com/matplotlib/matplotlib}{\texttt{matplotlib}}
\citep{matplotlib},
\href{https://github.com/numpy/numpy}{\texttt{NumPy}}
\citep{numpy1, numpy2},
\href{https://www.python.org/}{\texttt{Python}}
\citep{python},
\href{https://github.com/scipy/scipy}{\texttt{Scipy}}
\citep{scipy}.
\end{acknowledgements}

\bibliographystyle{aa}
\bibliography{MACS1149_H0}


\end{document}